%% file: Boron8Paper_29June2011.tex
\documentclass[aps,prl,twocolumn,groupedaddress,showpacs,amsmath,amssymb,floatfix,superscriptaddress]{revtex4}
\usepackage{graphicx}
\usepackage{times,color}
\usepackage{multirow}

\newcommand{\nuebar}{$\overline{\nu}_{e}$}

\newcommand{\BN}{$^{12}$B/$^{12}$N}

\newcommand{\flux}{$\times$10$^{6}$\,cm$^{-2}$s$^{-1}$}

\newcommand{\PhiB}{$\Phi_{^{8}\text{B}}$}

\newcommand{\PhiES}{$\Phi_{\text{ES}}$}

\newcommand{\spC}{$^{13}$C}

\newcommand{\spBe}{$^{11}$Be}
\newcommand{\spB}{$^8$B}
\newcommand{\spLi}{$^8$Li}
\newcommand{\bTl}{$^{208}$Tl}
\newcommand{\bTh}{$^{232}$Th}

\begin{document}

\title{Measurement of the \spB\ Solar Neutrino Flux with the KamLAND Liquid Scintillator Detector}
\include{KamLAND_Boron8Paper_Collaboration}

\begin{abstract}
  We report a measurement of the neutrino-electron elastic scattering rate from $^{8}$B solar neutrinos based on a 123 kton-day exposure of KamLAND.  The background-subtracted electron recoil rate, above a 5.5\,MeV analysis threshold is 1.49$\pm$0.14(stat)$\pm$0.17(syst) events per kton-day. Interpreted as due to a pure electron flavor flux with a \spB\ neutrino spectrum, this corresponds to a spectrum integrated flux of 2.77$\pm$0.26(stat)$\pm$0.32(syst)\,\flux.  The analysis threshold is driven by \bTl\ present in the liquid scintillator, and the main source of systematic uncertainty is due to background from cosmogenic \spBe. The measured rate is consistent with existing measurements and with Standard Solar Model predictions which include matter enhanced neutrino oscillation.
\end{abstract}


\maketitle
Neutrinos from \spB\ beta-decay (Q$\sim$18\,MeV) dominate the
high-energy portion of the solar neutrino spectrum and have been a
crucial source for solar neutrino experiments. Such experiments,
together with reactor anti-neutrino experiments, have established
flavor change through Large Mixing Angle\,(LMA) neutrino-oscillation
with matter effects, introduced by Mikheev, Smirnov and
Wolfenstein\,(MSW)~\cite{Mikheev:1986gs, Wolfenstein:1977ue}, as a
consistent resolution of the solar neutrino problem~\cite{PDG:2010,
  Abe:2008ee, Fukuda:2002pe, cravens:2008zn, Aharmim:2007nv,
  Aharmim:2005gt, Aharmim:2008kc, Aharmim:2009gd}.  Measurements of
neutrino-electron scattering using water-Cherenkov techniques have
been achieved in a number of different detectors:
Kamiokande~\cite{Hirata:1989zj, Hirata:1991ub},
Super-Kamiokande~\cite{Fukuda:2002pe, cravens:2008zn,Abe:2010hy} and
SNO~\cite{Aharmim:2007nv, Aharmim:2005gt, Aharmim:2008kc,
  Aharmim:2009gd}.  An advantage of this detection technique is
sensitivity to the neutrino direction of incidence, although analysis
thresholds of $\sim$5\,MeV are typically imposed. Recently SNO has
achieved a threshold of $\sim$3.5\,MeV \cite{Aharmim:2009gd}.  Liquid
scintillator detectors may also be used to measure neutrino-electron
elastic scattering; while insensitive to the neutrino direction, these
detectors have better energy resolution and have the possibility of
lower energy thresholds.  A measurement with a threshold of $\sim$
3.0\,MeV, the lowest to date, has been reported by Borexino~\cite{Bellini:2008mr}.  In this paper we report an independent,
liquid scintillator based measurement of the \spB\ flux with KamLAND.

The KamLAND detector consists of 1~kton of liquid scintillator (LS)
confined in a 6.5-m-radius, 135$\mu$m-thick, spherical balloon. The
LS is 80\% dodecane, 20\% pseudocumene, and 1.36$\pm$0.03\,g/l of PPO;
the density is 0.78\,g/cm$^{3}$. The balloon is suspended in purified
mineral oil within a 9-m-radius stainless steel sphere (SSS). The
scintillation light is recorded by an array of 1879 photomultiplier
tubes (PMTs) mounted on the inner surface of the SSS; 554 are reused
20-inch PMTs from Kamiokande, the remainder are new 20-inch PMTs with
the photo-cathode masked down to 17-inch. The 17-inch tubes have
better timing and single-photo-electron (spe) resolution.  A 3.2 kton,
water-Cherenkov detector surrounds the SSS. It is used as a muon
anti-coincidence counter and as shielding against external $\gamma$-rays and
neutrons.

Event position (vertex) and energy are reconstructed based on the
photon hit-time and charge distributions and a detector response model which is calibrated by
periodically deploying radioactive sources: $^{203}$Hg, $^{137}$Cs,
$^{68}$Ge, $^{65}$Zn, $^{60}$Co, $^{241}$Am$^{9}$Be,
$^{210}$Po$^{13}$C.  For the current analysis, the reconstruction
omits the 20-inch PMTs due to their poorer timing and spe resolution;
in this mode the energy resolution is 7\%/$\sqrt{\mathrm{E}(\mathrm{MeV})}$.  

The signal of interest, neutrino-electron elastic scattering from \spB\ solar
neutrinos, produces recoil electrons with reconstructed energy up to
$\sim$20\,MeV.  We construct a \textit{vertex}-$\chi^2$ test to select
electron-recoil-like events.  The test compares the observed PMT
charge and hit time distributions as a function of the distance to the event to the expected distributions.
The expected distributions and selection criteria were developed and calibrated
based on source calibration data and muon-tagged $\beta^-$events from
cosmogenic \BN\ . The efficiency of the selection is
$0.99\pm0.01$.
\begin{table}[tb]
  \caption{Production rate of muon-spallation isotopes that dominate the background (in events per kton-day)~\cite{Abe:2008spall}. \textit{Non-bright} muons produce less than 7$\times10^5$\,p.e. in the PMT array. }
\label{tab:spall}
\begin{tabular}{l|c|c|c|c}
\hline \hline
Isotope & Q[MeV] & $\tau_{1/2}$[s] & All Muons & Non-Bright Muons\\
\hline
\spLi & 16 & 0.84 & 15.6$\pm$3.2 & 0.7$\pm$0.4 \\ 
\spB & 18 & 0.77 & 10.7$\pm$2.9 & $<$0.006 \\
\spBe & 11.5 & 13.8 & 1.4$\pm$0.3 & 0.27$\pm$0.30 \\
\hline \hline
\end{tabular}
\end{table}

The analysis is carried out in \textit{reconstructed} energy with a
threshold of 5.5 MeV, which corresponds to a 5\,MeV threshold in physical
electron recoil energy. The threshold is chosen to reject background
from \bTl(Q$=$5.0\,MeV, $\tau_{1/2}$=3.05\,min) produced by residual \bTh\ 
contamination present in the LS at a level of $($7.90$\pm$0.25$)\times$10$^{-17}$\,g/g. 

An energy-scale model was developed to convert physical energy to
reconstructed energy. The model includes the linear response of the LS
and particle-dependent non-linearities such as scintillator quenching
and Cherenkov light production.  The energy scale is constrained by calibration data, tagged $\alpha$-particles from
$^{214}$Bi-$^{214}$Po/$^{212}$Bi-$^{212}$Po decays from residual
$^{238}$U/$^{232}$Th in the LS and high energy $\beta^-$ decays (Q
$\sim$14\,MeV) from cosmogenic \BN\ .

\begin{table}[tb]
\caption{Estimated background contributions for the full exposure and after all cuts. The shorter lived spallation products $^{12}$B, $^{12}$N, $^{9}$C, $^{9}$Li, and $^{8}$He are also considered and their contributions are summarized by ``Spallation Other". 
}
\label{tab:back}
\begin{tabular}{l c c c}
\hline \hline
Background & \multicolumn{3}{c}{Counts} \\
\hline
Spallation \spBe &89.1&$\pm$&19.1 \\
Spallation \spLi &20.5&$\pm$&4.0 \\
Spallation \spB &11.0&$\pm$&3.0 \\
Spallation Other & 0.4&$\pm$&0.6 \\
External gamma rays &25.2&$\pm$&12.6 \\
$^{8}$B CC on $^{13}$C GND & 5.8&$\pm$& 1.4 \\
Reactor \nuebar &1.6&$\pm$&0.1 \\
$^{8}$B CC on $^{13}$C 3.51\,MeV & 1.1&$\pm$&0.4 \\
$hep$ ES & 0.6&$\pm$&0.1 \\
Atmospheric $\nu$ &2.0&$\pm$&2.0 \\
\hline
Total & 157.3&$\pm$&23.6 \\
\hline \hline
\end{tabular}
\end{table}

The measurement reported here is restricted to data acquired before the
KamLAND LS purification. The data was taken between April 2002 through
April 2007, and corresponds to 1432.1 days of run-time.  To reduce the
external $\gamma$-ray background the fiducial volume (FV) is reduced to a
6-m-high, 3-m-radius cylinder centered at the detector center. This shape defines an optimal analysis volume that takes advantage of further shielding by the ultra-pure LS from backgrounds coming from the cavern's rock walls and a large steel deck which caps the experiment. The
FV fraction and systematic uncertainty is determined from
short-lived muon spallation products, mainly $^{12}$B(Q$=$13.4\,MeV,
$\tau_{1/2}=$20.2\,ms), which are assumed to be produced uniformly in the
LS.  The FV fraction is taken as the ratio of the number of
spallation products that reconstruct inside the analysis cylinder to
the number that reconstruct in the full LS volume.  This fraction
combined with the total LS volume in the balloon, which was measured
directly during construction to be 1171$\pm$25\,m$^{3}$, gives a
fiducial volume of 176.4\,m$^{3}$ with an uncertainty of
3.1\%.  When all selection cuts, to be described, are applied the remaining exposure is 123\,kton-days.

This paper focuses on neutrino-electron elastic scattering,
\mbox{$\nu+e^{-}\rightarrow\nu+e^{-}$} (ES) due to \spB\ solar
neutrinos.  With 3.423$\times$10$^{32}$\,$e^{-}$ targets per kton of LS, we
expect, including oscillation, 1.19 recoil events per kton-day
from \spB\ in the energy-region of interest (ROI) 5.5--20\,MeV.  In
our calculation we adopt the standard solar model (SSM) of Serenelli
et al.~\cite{Serenelli:2009yc} which uses the solar abundances of Asplund et
al. \cite{Asplund:2009fu} (AGSS09). This predicts a total \spB\
neutrino flux, independent of neutrino flavor, of
\PhiB\,=\,$4.85_{-0.58}^{+0.58}$\flux.  We use the ES cross section
from ~\cite{PhysRevD.51.6146}; the \spB-$\nu_{e}$ spectrum of Winter
et al.~\cite{Winter:2004kf}; the oscillation parameters from the
global analysis analysis in~\cite{Abe:2008ee}; and fold in the
detector response.  ES from $hep$ neutrinos and neutrino interactions on
carbon in the LS are treated as a background and are discussed later in the paper.  We note that the best choice of solar abundances to use in the SSM is an
unresolved question.  In AGSS09 the helioseismic discrepancy which
emerged when the abundances of Asplund et
al.\cite{Asplund:2005}(AGS05) were adopted still persists, although
it is not as severe.  If the older, 1-dimensional, solar abundance
evaluation of Grevesse and Sauval~\cite{Grevesse:1998bj}(GS98) is
used, the SSM has excellent consistency with helioseismology and the
predicted \spB\ flux is \PhiB\,=\,$5.88_{-0.65}^{+0.65}$\flux. Assuming no
sterile neutrinos, the total \spB\ flux is directly observed in the
SNO neutral current measurement to be
\PhiB\,=\,$5.140_{-0.207}^{+0.197}$\flux~\cite{Aharmim:2009gd}.

With the 5.5\,MeV analysis threshold the background is dominated by
decays of light isotopes produced by muon spallation.  An in-depth
study of muon activation at KamLAND can be found in~\cite{Abe:2008spall}.  A study of light isotope production shows that most ($>\,80\%$) light isotope backgrounds are correlated with muons which produce more than 700,000\, photo-electrons (p.e.) in the PMT array. We denote these as \emph{bright} muons. The rate of bright muons is 0.037\,Hz, while the total rate of muons passing through the LS is 0.198$\pm$0.014\,Hz. We
apply a number of muon-related cuts to reduce these spallation
backgrounds. All events within 200\,ms of a preceding muon are
rejected. This time veto of the full detector significantly reduces
the background from \BN.  For non-bright muons, for which the muon
tracking algorithm converged successfully ---well-tracked muons--- a
5\,s veto is applied within a 3-m-radius cylinder around the muon track. Using the \BN\, candidates we determine that 97$\pm$6\% of spallation products are contained within this cylinder. For bright muons and any muon with a poorly reconstructed track, the
5\,s veto is applied to the full detector.  These cuts reduce the
exposure by 62.4$\pm$0.1\%, to a total exposure of 123
kton-days. The spallation background events remaining after these cuts are
expected to come mainly from long lived ($\tau >1\,$s) spallation
products, \spBe, \spLi, \spB. The total production rates of these key
isotopes~\cite{Abe:2008spall}, along with the rates correlated to non-bright muons, are summarized in
Table~\ref{tab:spall}.
\begin{figure}
\includegraphics{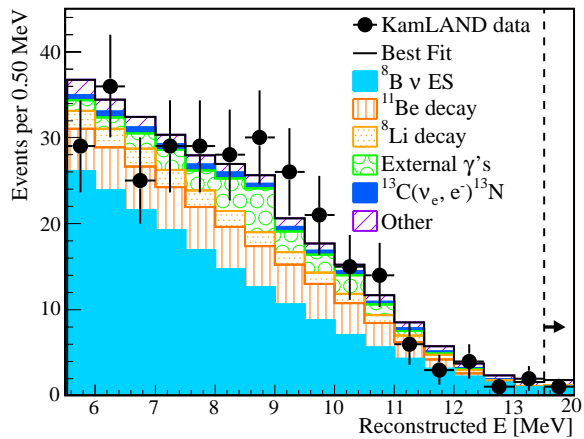}
\caption[] {Energy spectrum of \spB\ candidates with the best-fit spectrum and background components from the unbinned energy and rate analysis. The histograms display the results in bins of 0.5\,MeV except for the last bin which due to limited statistics extends from 13.5\,MeV to 20\,MeV.}
\label{fig:nuEnergy}
\end{figure}

The next-largest background is from external gamma rays which
are primarily the result of ($n$,$\gamma$) reactions in the surrounding rock cavern and
stainless steel detector elements. The external $\gamma$-ray spectrum has peaks in reconstructed energy
at 8.5\,MeV and 10\,MeV from stainless steel, and at 5.5\,MeV from
neutron capture on silicon in the rock.  The cylindrical fiducial
volume was chosen to optimize the shielding for a given exposure.  The
closest points to the cylindrical external rock cavity are at the
balloon equator, while the closest stainless steel component is a
balloon support structure at the top of the detector.  A GEANT4-based
Monte Carlo~\cite{Agostinelli:2002hh, Allison:2006ve}, including the
full detector and shielding geometry, simulated the effect of LS
self-shielding from sources in the stainless steel and the rock.  The
simulation indicates that gammas are attenuated approximately
exponentially with an attenuation length of 53.0$\pm$0.1\,cm for
rock and 50.7$\pm$0.1\,cm for stainless steel.  Using data within a
6-m-radius volume, we observe attenuation lengths for gammas from the rock and
stainless steel of 59.0$\pm$1.9\,cm and 54.9$\pm$1.9\,cm respectively.
We estimate 25.2$\pm$12.6 electron-recoil-like
events from external $\gamma$-rays in the ROI within the cylindrical
fiducial volume. The uncertainty in the estimate comes from the
difference in the observed and simulated attenuation lengths.
\begin{table}
  \caption{The systematic uncertainties associated with the unbinned fit to the energy spectrum of the \spB\ candidates. The detection efficiency is dominated by our fiducial volume uncertainty.}
\begin{center}
\begin{tabular}{ c c c }
\hline
Source & Uncertainty (\%) \\
\hline
\spBe & 10.8 \\
\spLi\ and \spB & 3.3 \\
External gamma rays & 6.8 \\
Other Backgrounds &  1.1 \\
Detection Efficiency & 6.3 \\
Energy Scale & 0.8 \\
\hline
\hline
\end{tabular}
\end{center}
\label{table:syst}
\end{table}

\begin{figure*}
\includegraphics{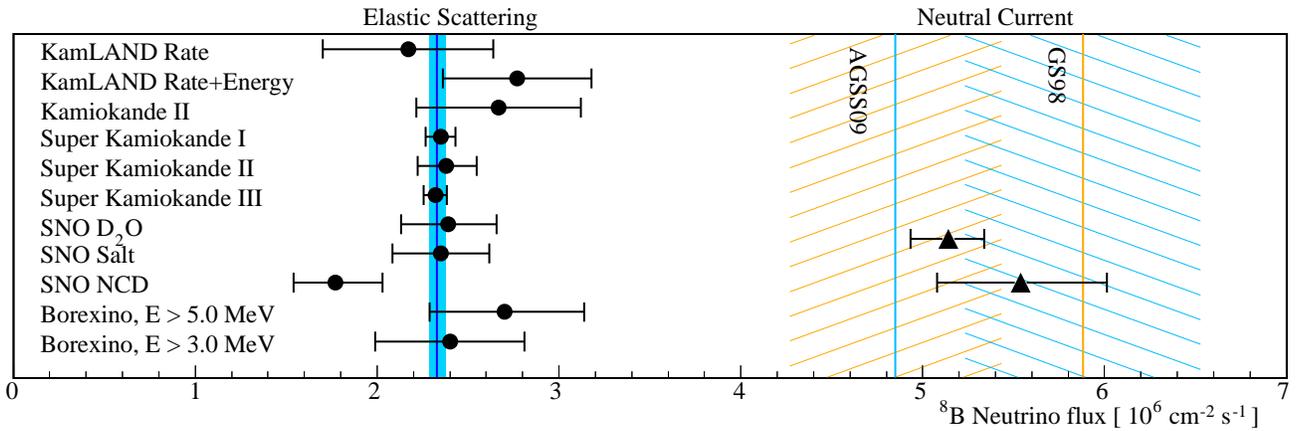}
\caption{Summary of measurements of the \spB\ flux using
  neutrino-electron scattering: this work KamLAND, Kamiokande II \cite{Hirata:1989zj, Hirata:1991ub}, Super Kamiokande I \cite{Fukuda:2002pe}, Super Kamiokande II \cite{cravens:2008zn}, Super Kamiokande III \cite{Abe:2010hy}, SNO D$_{2}$O \cite{Aharmim:2007nv, Aharmim:2009gd}, SNO Salt \cite{Aharmim:2005gt, Aharmim:2009gd}, SNO NCD \cite{Aharmim:2008kc}, and Borexino \cite{Bellini:2008mr}.  The SNO neutral current measurements \cite{Aharmim:2009gd,Aharmim:2008kc} are shown for reference with closed triangles.}
\label{fig:expSum}
\end{figure*}

As remarked earlier, events from $hep$ solar neutrinos and solar
neutrino interactions on carbon are treated as a background. Using the
SSM with AGSS09 and the $hep$ spectrum from~\cite{Bahcall:1987jc}, we
estimate, including oscillation, 0.6$\pm$
0.1 electron recoil events from $hep$ neutrinos in the ROI.
In our calculation we use the ES cross section, neutrino oscillation
parameters and detector response as was done for the \spB\ ES calculation.  The uncertainty is dominated by the difference in the flux prediction of the SSM
with the AGSS09 versus with the GS98 solar abundances.  

There are 4.30$\times$10$^{31}$ carbon nuclei per kton of LS if we assume a natural \spC\ of 1.10\%. Using the
cross sections calculated in~\cite{Fukugita:1989wv}, we find the largest
$\nu_{e}$-C scattering background contribution to be from charged
current (CC) scattering,\mbox{$^{13}$C$+\nu_e\rightarrow^{13}$N$+e^{-}$} from
\spB. We estimate, including oscillation, that scattering to the
ground state of $^{13}$N produces 5.8$\pm$1.4 events in the ROI; and scattering to the 3.51\,MeV
first excited, which decays by proton emission, contributes
1.1$\pm$ 0.4 electron-recoil-like events in
the ROI.  In this estimate an additional uncertainty of 30\% on the
cross section is included.  The contribution from higher states of
$^{13}$N, neutral current (NC) scattering by \spB-$\nu$ and $hep-\nu$ NC and CC interactions
on carbon is estimated to be less than $0.13$ events in the ROI, and
is considered negligible given the other backgrounds. These calculations use the
same solar-$\nu$ fluxes, spectra, detector response and
oscillation parameters as before.  The effect of using the oscillation parameters from the global analyses in~\cite{Gando:2010aa,Abe:2010hy} were investigated but do not change the result due to the large flux and cross section uncertainties.

In addition to these neutrino interactions, we expect a small
background from atmospheric neutrino interactions. Assuming a flat
spectrum, we use the candidate events in the range of 20\,MeV to
35\,MeV to extrapolate the atmospheric neutrino contribution in the
signal region, and find 2.0$\pm$2.0 events.

A background from positrons from inverse beta decay induced by reactor
antineutrinos is tagged by coincidence with delayed neutron capture $
\gamma$-rays.  The mean capture time in KamLAND is
$\sim$207\,$\mu$s. The tagging cuts are: a delayed energy cut,
2.04\,MeV$<$E$<$2.82\,MeV, to select the 2.2\,MeV n-p capture
$\gamma$; a timing cut, 0.5$\mu$s$<$$\Delta$T$<$660$\mu$s; and a
position cut, $\Delta$R$<$1.6m.  $\Delta$T and $\Delta$R refer
respectively to the time and distance between the prompt positron and
delayed capture $\gamma$.  The tagging efficiency is
0.940$\pm$0.006.  The reactor antineutrino flux at KamLAND is
dominated by Japanese power reactors, we calculate the flux and
spectrum using data provided by the Japanese power companies and a
simplified reactor model ~\cite{Nakajima:2006re}. The residual
background is 1.6$\pm$0.1, where the uncertainty in the neutrino oscillation parameters is included.
 
The total background, broken down in Table~\ref{tab:back}, is
estimated to be 157.3$\pm$23.6 events.  While
\spLi\ has the largest production rate, \spBe, due to its long
half-life ($\tau _{1/2}=13.8$\,s), is the largest contribution to the
total background.

After applying all the cuts, 299 electron recoil
candidates are found in the ROI in the
123 kton-day exposure. Subtracting the
157.3$\pm$23.6 background events gives a \spB\
rate of 1.16$\pm$0.14(stat)$\pm$0.21(syst)\,events per kton-day which is consistent with our prediction including neutrino oscillation. If we neglect neutrino oscillation, assume a pure $\nu_e$ flux,  and correct for the 5.5\,MeV threshold, the measured rate corresponds to a spectrum integrated flux of \PhiES\,=\,2.17$\pm$0.26(stat)$\pm$0.39(syst)\,\flux. 

A second analysis of this data uses an unbinned maximum likelihood fit to the \spB\ candidate energy
spectrum. The normalization of the individual backgrounds are allowed
to vary but are constrained by a penalty to the expected value. The
best-fit rate is 1.49$\pm$0.14(stat)$\pm$0.17(syst)\,events per kton-day, with a
goodness-of-fit of 49\%. Once again assuming a pure $\nu_e$ flux, this corresponds to a spectrum integrated flux of \PhiES\,=\,2.77$\pm$0.26(stat)$\pm$0.32(syst)\,\flux. Fig.~\ref{fig:nuEnergy} shows the energy spectrum of the candidate events with the best-fit solar neutrino and background spectra. This analysis does not include
the small change in the shape of the neutrino spectra due to
oscillation. Including this effect does not significantly change
the best-fit rate, \PhiES\,=\, 2.74$\pm$0.26(stat)$\pm$0.32(syst)\,\flux, with a
goodness of fit of 57\%.

The KamLAND result is compared to other measurements of the total
\spB\ flux through neutrino-electron elastic scattering in
Fig.~\ref{fig:expSum}. For this comparison the reported statistical
and systematic errors are added in quadrature. The mean of the
experiments, weighted by their uncertainties, is
\PhiES\,=\,2.33$\pm$0.05\,\flux\,. This is dominated by the very
precise Super-Kamiokande water-Cherenkov measurement. The reduced
$\chi^{2}/$NDF is 6.6/ 8 where the KamLAND Rate $+$ Energy result and the Borexino E$>$3\,MeV result are used in the calculation.

\begin{figure}
\includegraphics{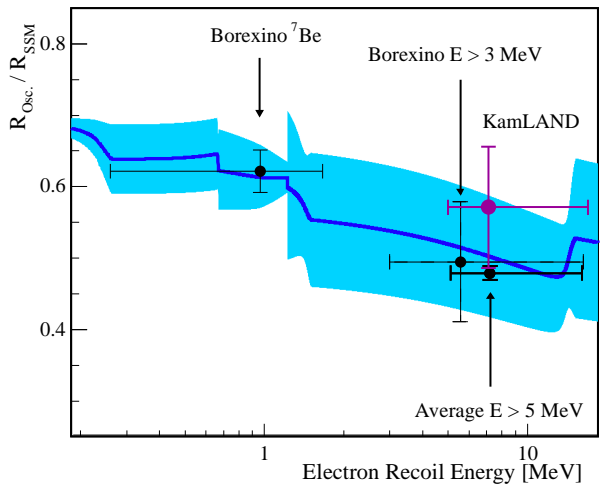}
\caption[] {Ratio of the measured (threshold corrected) solar
  neutrino-electron elastic scattering rate and that predicted by the AGSS09-SSM as a
  function of energy (data points). The average of the water-Cherenkov detector results above 5\,MeV is shown.  The horizontal error bars indicate the range of energy over which the flux was measured. The central value indicates the flux averaged energy for that range. The solid curve (and $\pm\,1\,\sigma$ band) shows the ratio of AGSS09 with neutrino oscillation included to the prediction without oscillation. The discontinuities in the prediction arise as different neutrino sources become significant, the large uncertainty in the CNO neutrinos is evident at $\sim$1.2\,MeV.}
\label{fig:ratio}
\end{figure}

A key prediction of the LMA-MSW solution to the solar neutrino problem is
the expected transition from matter-dominated to vacuum-dominated
oscillations, depending on the neutrino energy and the region of the
solar interior probed by the neutrinos as they journey out of the sun.
A general feature is that the survival probability, defined as the
ratio of the oscillated $\nu_{e}$-flux to the unoscillated flux, tends
to increase with decreasing solar neutrino energy. This effect is the result of neutrino species ($^{8}$B, $^{7}$Be, $pp$, etc.)
either moving off, or being produced in a region not satisfying the MSW resonance
condition. These neutrinos are then described by vacuum oscillation. The curve in
Fig.~\ref{fig:ratio} shows the ratio of the expected
neutrino-electron elastic scattering rates with and without
oscillation calculated using the AGSS09-SSM.  The data points are the
survival probabilities deduced, relative to the SSM prediction, from
neutrino-electron elastic scattering measurements at Borexino, KamLAND
and the combined water-Cherenkov experiments.  The Borexino \spB\
measurement with a 3.0\,MeV threshold~\cite{Bellini:2008mr} and the
$^{7}$Be flux measurement~\cite{Bellini:2011rx, Arpesella:2008mt} are
consistent with LMA-MSW prediction.  The measurements are not yet
precise enough to resolve the issue of solar abundances.

This letter reports a measurement of the \spB\ solar neutrino flux
measured through neutrino-electron elastic scattering in the KamLAND
LS.  Our result is in good agreement with existing measurements from both
water-Cherenkov and liquid-scintillator-based detectors.  In the
data-set presented, \bTl\ background limits the KamLAND analysis
threshold to 5.5\,MeV. We expect to achieve a lower threshold in a
future analysis of data taken after purification of the KamLAND LS.

The KamLAND experiment is supported by the Grant-in-Aid for Specially
Promoted Research under grant 16002002 of the Japanese Ministry of
Education, Culture, Sports, Science and Technology; the World Premier
International Research Center Initiative (WPI Initiative), MEXT,
Japan; and under the U.S.\ Department of Energy (DOE) Grants
DE-FG03-00ER41138, DE-AC02-05CH11231, and DE-FG02-01ER41166, as well
as other DOE grants to individual institutions.  The reactor data are
provided by courtesy of the following electric associations in Japan:
Hokkaido, Tohoku, Tokyo, Hokuriku, Chubu, Kansai, Chugoku, Shikoku,
and Kyushu Electric Power Companies, Japan Atomic Power Co., and Japan
Atomic Energy Agency.  The Kamioka Mining and Smelting Company has
provided service for activities in the mine.

\bibliography{Boron8Paper_29June2011}

\end{document}

%% file: KamLAND_Boron8Paper_Collaboration.tex
%
%
%

\newcommand{\tohoku}{\affiliation{Research Center for Neutrino
    Science, Tohoku University, Sendai 980-8578, Japan}}
\newcommand{\alabama}{\affiliation{Department of Physics and
    Astronomy, University of Alabama, Tuscaloosa, Alabama 35487, USA}}
\newcommand{\lbl}{\affiliation{Physics Department, University of
    California, Berkeley and \\ Lawrence Berkeley National Laboratory, 
Berkeley, California 94720, USA}}
\newcommand{\caltech}{\affiliation{W.~K.~Kellogg Radiation Laboratory,
    California Institute of Technology, Pasadena, California 91125, USA}}
\newcommand{\colostate}{\affiliation{Department of Physics, Colorado
    State University, Fort Collins, Colorado 80523, USA}}
\newcommand{\drexel}{\affiliation{Physics Department, Drexel
    University, Philadelphia, Pennsylvania 19104, USA}}
\newcommand{\hawaii}{\affiliation{Department of Physics and Astronomy,
    University of Hawaii at Manoa, Honolulu, Hawaii 96822, USA}}
\newcommand{\kansas}{\affiliation{Department of Physics,
    Kansas State University, Manhattan, Kansas 66506, USA}}
\newcommand{\lsu}{\affiliation{Department of Physics and Astronomy,
    Louisiana State University, Baton Rouge, Louisiana 70803, USA}}
\newcommand{\stanford}{\affiliation{Physics Department, Stanford
    University, Stanford, California 94305, USA}}
\newcommand{\ut}{\affiliation{Department of Physics and
    Astronomy, University of Tennessee, Knoxville, Tennessee 37996, USA}}
\newcommand{\tunl}{\affiliation{Triangle Universities Nuclear
    Laboratory, Durham, North Carolina 27708, USA and \\
Physics Departments at Duke University, North Carolina Central University,
and the University of North Carolina at Chapel Hill}}
\newcommand{\wisc}{\affiliation{Department of Physics, University
    of Wisconsin, Madison, Wisconsin 53706, USA}}  
\newcommand{\cnrs}{\affiliation{CEN Bordeaux-Gradignan, IN2P3-CNRS and
    University Bordeaux I, F-33175 Gradignan Cedex, France}}
\newcommand{\ipmu}{\affiliation{Institute for the Physics and Mathematics of the 
    Universe, Tokyo University, Kashiwa 277-8568, Japan}}
\newcommand{\nikhef}{\affiliation{Nikhef, Science Park, Amsterdam, Netherlands}}

    
\newcommand{\atlanlnow}{\altaffiliation{Present address: Subatomic Physics Group, Los Alamos National Laboratory, Los Alamos, NM 87545, USA}}
    
\newcommand{\atksunow}{\altaffiliation{Present address: Department of Physics, Kansas State University, Manhattan, Kansas 66506, USA}}

\newcommand{\attokyotechnow}{\altaffiliation{Present address: Department of Physics, Tokyo Institute of Technology, Tokyo 152-8551, Japan}}

\newcommand{\atregisnow}{\altaffiliation{Present address: Department of Physics and Computational Science, Regis University, Denver, Colorado 80221, USA}}

\newcommand{\atfnalnow}{\altaffiliation{Present address: Fermi National Accelerator Laboratory, Batavia, Illinois 60510, USA}}

\newcommand{\atsnolabnow}{\altaffiliation{Present address: SNOLAB, Lively, ON P3Y 1N2, Canada}}

\newcommand{\atllnlnow}{\altaffiliation{Present address: Lawrence Livermore National Laboratory, Livermore, California 94550, USA}}

\newcommand{\atucdnow}{\altaffiliation{Present address: Department of Physics, University of California, Davis, California 95616, USA}}

\newcommand{\atuwnow}{\altaffiliation{Present address:  CENPA, University of Washington, Seattle, Washington 98195, USA}}

\newcommand{\atumdnow}{\altaffiliation{Present address: Department of Physics, University of Maryland, College Park, Maryland 20742, USA}}

\newcommand{\atmitnow}{\altaffiliation{Present address: Department of Physics, Massachusetts Institute of Technology, Cambridge, MA 02139, USA}}

\newcommand{\atrowannow}{\altaffiliation{Present address: Department of Physics and Astronomy, Rowan University, Glassboro, New Jersey 08028, USA}}

\newcommand{\atsdnow}{\altaffiliation{Present address: Department of Physics, University of South Dakota, South Dakota 57069, USA}}

\newcommand{\atlsunow}{\altaffiliation{Present address: Department of Physics, Louisiana State University, Baton Rouge, LA 70803 USA}}

\newcommand{\atuwjnow}{\altaffiliation{Jointly at: Center for Experimental Nuclear Physics and Astrophysics, University of Washington, Seattle, Washington 98195, USA}}

%
%
\author{S.~Abe}\tohoku
\author{K.~Furuno}\tohoku
\author{A.~Gando}\tohoku
\author{Y.~Gando}\tohoku
\author{K.~Ichimura}\tohoku
\author{H.~Ikeda}\tohoku
\author{K.~Inoue}\tohoku\ipmu
\author{Y.~Kibe}\attokyotechnow\tohoku
\author{W.~Kimura}\tohoku
\author{Y.~Kishimoto}\tohoku
\author{M.~Koga}\tohoku\ipmu
\author{Y.~Minekawa}\tohoku
\author{T.~Mitsui}\tohoku
\author{T.~Morikawa}\tohoku
\author{N.~Nagai}\tohoku
\author{K.~Nakajima}\tohoku
\author{K.~Nakamura}\tohoku\ipmu
\author{M.~Nakamura}\tohoku
\author{K.~Narita}\tohoku
\author{I.~Shimizu}\tohoku
\author{Y.~Shimizu}\tohoku
\author{J.~Shirai}\tohoku
\author{F.~Suekane}\tohoku
\author{A.~Suzuki}\tohoku
\author{H.~Takahashi}\tohoku
\author{N.~Takahashi}\tohoku
\author{Y.~Takemoto}\tohoku
\author{K.~Tamae}\tohoku
\author{H.~Watanabe}\tohoku
\author{B.D.~Xu}\tohoku
\author{H.~Yabumoto}\tohoku
\author{E.~Yonezawa}\tohoku
\author{H.~Yoshida}\tohoku
\author{S.~Yoshida}\tohoku
%
\author{S.~Enomoto}\atuwjnow\ipmu
\author{A.~Kozlov}\ipmu
\author{H.~Murayama}\ipmu\lbl
%
\author{C.~Grant}\alabama
\author{G.~Keefer}\atllnlnow\alabama
\author{D.~McKee}\atksunow\alabama
\author{A.~Piepke}\ipmu\alabama
%
\author{T.I.~Banks}\lbl
\author{T.~Bloxham}\lbl
\author{J.A.~Detwiler}\lbl
\author{S.J.~Freedman}\ipmu\lbl
\author{B.K.~Fujikawa}\ipmu\lbl
\author{K.~Han}\lbl
\author{R.~Kadel}\lbl
\author{T.~O'Donnell}\lbl
\author{H.M.~Steiner}\lbl
\author{L.A.~Winslow}\atmitnow\lbl
%
\author{D.A.~Dwyer}\caltech
\author{C.~Mauger}\atlanlnow\caltech
\author{R.D.~McKeown}\caltech
\author{C.~Zhang}\caltech
%
\author{B.E.~Berger}\colostate
%
\author{C.E.~Lane}\drexel
\author{J.~Maricic}\drexel
\author{T.~Miletic}\atrowannow\drexel
%
\author{M.~Batygov}\atsnolabnow\hawaii
\author{J.G.~Learned}\hawaii
\author{S.~Matsuno}\hawaii
\author{S.~Pakvasa}\hawaii
\author{M.~Sakai}\hawaii
%
\author{G.A.~Horton-Smith}\ipmu\kansas
\author{A.~Tang}\kansas
%
\author{K.E.~Downum}\stanford
\author{G.~Gratta}\stanford
\author{K.~Tolich}\stanford
%
\author{Y.~Efremenko}\ipmu\ut
\author{Y.~Kamyshkov}\ut
\author{O.~Perevozchikov}\atlsunow\ut
%
\author{H.J.~Karwowski}\tunl
\author{D.M.~Markoff}\tunl
\author{W.~Tornow}\tunl
%
\author{K.M.~Heeger}\ipmu\wisc 
%
\author{F.~Piquemal}\cnrs
\author{J.-S.~Ricol}\cnrs
%
\author{M.P.~Decowski}\ipmu\lbl\nikhef

\collaboration{The KamLAND Collaboration}\noaffiliation